\documentclass[showpacs,amsmath,amssymb,showkeys,pre]{revtex4}
\usepackage{graphicx}

\newcommand{\beq}{\begin{equation}}
\newcommand{\eeq}{\end{equation}}
\newcommand{\beqa}{\begin{eqnarray}}
\newcommand{\eeqa}{\end{eqnarray}}
\newcommand{\non}{\nonumber}
\newcommand{\s}{\sigma}
\newcommand{\JJ}{J_{i_1\dots i_p}}
\newcommand{\KK}{K_{l_1\dots l_r}}
\newcommand{\e}{\epsilon}

\begin{document}

\title{Coexistence of supersymmetric and supersymmetry-breaking states \\ in spherical
spin-glasses}

\author{Alessia Annibale$^{\dagger \ddagger}$, Giulia Gualdi$^{\dagger}$, Andrea Cavagna$^{\dagger \ddagger}$}

\affiliation{$^\dagger$ Dipartimento di Fisica, Universit\`a di Roma ``La Sapienza'', Piazzale Aldo Moro 5, 00185 Roma, 
Italy}

\affiliation{$^\ddagger$ Center for Statistical Mechanics and Complexity, INFM Roma ``La Sapienza''}

\date{June 21, 2004}

\begin{abstract}
The structure of states of the perturbed $p$-spin spherical spin-glass is analyzed.
At low enough free energy metastable states have a supersymmetric structure, while at higher 
free energies the supersymmetry is broken. The transition between the supersymmetric and the 
supersymmetry-breaking phase is triggered by a change in the stability of states. 
\end{abstract}


\maketitle

It is well known that mean-field spin-glasses can be divided into two broad categories. On the one hand there are
systems with a low temperature thermodynamic phase described by full (or continuous) breaking of the replica symmetry 
(FRSB). In these systems the static transition temperature $T_s$ coincides with the temperature $T_d$ where the
relaxation time diverges. Moreover, the dynamical correlation function exhibits single-step relaxation, either 
exponential (high temperature), or power-law (low temperature). The paradigm of these spin-glasses is the
Sherrington-Kirkpatrick (SK) model \cite{sk}. On the other hand, there are systems where the static and dynamic transitions
occur at different temperatures, namely $T_s < T_d$. In these systems thermodynamics is described by a one step
breaking of the replica symmetry (1RSB), and the dynamical correlation function develops, at low temperatures, a two-steps
relaxation, typical of structural glasses. The main representative of this second class is the $p$-spin spherical model
\cite{pspin}. In mean-field spin-glasses, it seems that a given type of static
behaviour is always accompanied by a given type of dynamical behaviour. No mixed cases are known.

In disordered systems statics and dynamics are linked to each other by the structure of states: lowest lying states,
i.e. ground states, dominate the statics, whereas the dynamics is ruled by metastable states with higher
free energy. It is therefore reasonable that the origin of the two different behaviours described above, and thus
of the connection between statics and dynamics, has to be sought in the structures of states of different systems.
It is common wisdom, for example, that metastable states play a much greater role in 1RSB than FRSB spin-glasses, 
and that for this reason dynamics is so much different from thermodynamics in 1RSB systems. However, it would be 
important to have a mathematical formalization of the different structure of states in the two classes.
Recent studies \cite{juanpe,noi1,noi2,noi3,leuzzi1,leuzzi2,leuzzi4, abm,noi4,gold} 
suggest that the Becchi-Rouet-Stora-Tyutin (BRST) supersymmetry \cite{brs,tito,ps1,ps2,zinn}
may provide such a formalization.

The BRST supersymmetry is a transformation which leaves invariant the action which one has to define in order to study
the structure of metastable states in spin-glasses \cite{jorge1,juanpe,noi1}, and in particular when the complexity, i.e.
the entropy of the states, is computed. As always in physics, the symmetry of the action may or may not be shared
by the actual physical states. More specifically, in spin-glasses the complexity is equal to the action evaluated in 
its stationary point: there are systems where the physical (stable) stationary point is symmetric, and other systems where 
it is not. In this last case the supersymmetry is spontaneously broken. After one year of debate, it seems now clear
that the structure of states of the SK model breaks the supersymmetry \cite{abm,noi4}, while in the $p$-spin spherical 
model the supersymmetry gives the correct set of physical states \cite{juanpe,leuzzi4}. This fact is in agreement with the other
known physical features of these systems. Supersymmetry breaking occurs because metastable states are marginal and very
unstable under external perturbations \cite{abm,noi4,gold}, 
whereas well shaped states, with finite second derivative and large 
barriers surrounding them, preserve the supersymmetry. This is exactly what we expect from the different dynamical behaviours 
and role of metastable states, in the SK and $p$-spin models.

Classifying spin-glasses according to whether or not
the supersymmetry is spontaneously broken seems therefore a sound criterion to 
distinguish different structures of states, and thus different static and dynamic behaviours. 
In order to understand
the physical differences of these two cases in a deeper way, it would be desirable to find a model where
both structures of states are present, and to study its analogies with the two archetypes  
described above, the SK and the $p$-spin model. In this paper we introduce such a model and show that supersymmetric and 
supersymmetry-breaking states can coexist in the same system, albeit at different free energies. We will argue that 
such a situation may be rather general in spin-glasses, and that the SK and $p$-spin models correspond to the
two extreme cases: the transition from supersymmetric to supersymmetry-breaking states takes place at the ground state in the 
SK model, whereas the transition occurs at the highest free energy in the $p$-spin spherical model.

Before starting our analysis, let us make a general remark. 
There is a dynamical form of the supersymmetry, which arises when studying the
Langevin dynamics of statistical models \cite{zinn,jorge2}. The important difference with its static 
counterpart described here, is that the dynamical supersymmetry is always broken at low
temperatures, in {\it all} models. In other words, it is an intrinsic consequence of ergodicity breaking,
linked to the break-down of the dynamical fluctuation-dissipation theorem \cite{jorge2}. 
On the other hand, the static supersymmetry we are discussing may or may not be broken, according to 
the particular structure of states displayed by each given model.

The Hamiltonian of the model we want to study is given by,
\beq
H=-\sum_{i_1<\dots <i_p}^N \JJ\ \s_{i_1} \dots \s_{i_p} - \ \
\e \sum_{l_1<\dots <l_r}^N \KK\ \s_{l_1} \dots \s_{l_r} \ ,
\label{H}
\eeq
where the spins $\s_i$ are real variables subject to the spherical constraint $\sum_i \s_i^2=N$,
and the Gaussian random couplings $\JJ$ and $\KK$ have variance $p!/2N^{p-1}$ and $r!/2N^{r-1}$,
respectively. We are therefore perturbing the $p$-spin spherical 
model with an extra $r$-body interaction term. It is known that $p+r$ spherical models may display 
a nontrivial thermodynamic behaviour when $p\geq 3$ and $r=2$: in that case there is a transition between
a 1RSB thermodynamic phase (low $\e$), to a FRSB phase (large $\e$) \cite{theo}. On the contrary, if
both $p$ and $r$ are strictly larger than two, we expect the model to have a normal 1RSB 
thermodynamic behaviour. This is the case we will analyze. What we will show is that, even in this apparently 
simple case, as soon 
as $\e\neq 0$ the structure of metastable states becomes much richer than the unperturbed 
$p$-spin model. In the following, all plots are done for $p=3$ and $r=4$. 

Metastable states are the local minima of the mean-field free energy, that is the Thouless-Anderson-Palmer (TAP)
free energy \cite{tap}. The first thing to do is therefore to work out the TAP free energy for this model. 
This can be done by following
the method used by Rieger in \cite{rieger}. We will work with  $\e\ll 1$, and thus keep only the lowest
order terms in $\e$. The TAP free energy density is \cite{giulia},
\beqa
\beta f_{TAP} = 
  &-&\frac{\beta}{N}\sum_{i_1<\dots <i_p}^N \JJ\ m_{i_1} \dots m_{i_p} - \ \
\e \frac{\beta}{N}\sum_{l_1<\dots <l_r}^N \KK\ m_{l_1} \dots m_{l_r}
-\ \frac{1}{2}\log(1-q)
\non \\
&-& \ \frac{\beta^2}{4}\left[ (p-1)q^p -pq^{p-1}+1\right]
\ - \  \e^2 \frac{\beta^2}{4}\left[ (r-1)q^r -rq^{r-1}+1 \right] \ ,
\eeqa
where $m_i=\langle \s_i\rangle$ are the local magnetizations, and $q$ is the self-overlap 
of a state, $q=\sum_i m_i^2/N$.
We had to keep the $O(\e^2)$ term since, after averaging  over $\KK$, also the linear term 
will become $O(\e^2)$. To compute the TAP complexity we closely follow the standard method, as
can be found, for example, in \cite{noi1}. The number of TAP states with free energy density $f$ is,
\beq
{\cal N}(f) =
\int \prod_i dm_i\ \delta[\partial_i \beta f_{TAP}(m)]  \ 
\det [\partial_i \partial_j \beta f_{TAP}(m)] \ 
\delta[\beta f_{TAP}(m)-\beta f]
\ ,
\label{ghirri}
\eeq
where, as usual, the modulus of the determinant has been dropped (for a discussion of this point 
see \cite{jorge1, ps1, ps2} and \cite{noi1}). 
We give an exponential representation for the two delta functions introducing the Lagrange multipliers 
$x_i$ and $u$, and of the determinant, with the anti-commuting Grassmann vectors $\{\bar\psi_i, \psi_i\}$ 
\cite{noi1}.  In this way we can write,
\beq
{\cal N}(f) =
  \int {\cal D}m\, {\cal D}x \,{\cal D}\bar\psi\, {\cal D}\psi\, du \ \
e^{S(m,x,\bar\psi,\psi,u;\,f)} \ ,
\label{int2}
\eeq
where the action $S$ is given by,
\beq
S(m,x,\bar\psi,\psi,u;\,f)= \beta \sum_i^N x_i \partial_i f_{TAP}(m) + \beta \sum_{ij}^N\bar\psi_i  \psi_j  
\partial_i \partial_j f_{TAP}(m) + \beta u f_{TAP}(m)-\beta u f \ .
\label{action}
\eeq
We work at the annealed level, which is exact for $\e=0$. After averaging  the number ${\cal N}(f)$
over the disorder $\JJ$ 
and $\KK$, the integrals in ${\cal D}m\, {\cal D}x \,{\cal D}\bar\psi\, 
{\cal D}\psi$ can be performed exactly, provided that we introduce the parameters,
\beq
T = \frac1N \sum_i^N \bar\psi_i \,\psi_i 
\quad\quad\quad
B = \frac1N \sum_i^N x_i\,  m_i 
\quad\quad\quad
q = \frac1N \sum_i^N m_i \, m_i \ .
\label{vara}
\eeq
After some algebra we obtain,
\beq
\overline{{\cal N}(f)} =
  \int  dT\, dB\, dq\,  du \  
e^{N \hat S(T,B,q,u;\,f)} \ ,
\label{zizu}
\eeq
where the effective action $\hat S$ given by \cite{giulia}, 
\beqa
\hat S &=& 
\beta u \left[\, g(q) +\e^2 h(q) - f\,\right] 
+ \left(B^2-T^2\right)\left[ \frac14 p(p-1)\beta^2 q^{p-2} +\e^2 \frac14 r(r-1)\beta^2 q^{r-2} \right]
\non \\
&-& \frac12\log\left(\frac12\beta^2 p q^{p-2}+\frac12 \e^2 \beta^2 r q^{r-2}\right) 
-\log T
+ \frac14\beta^2 u^2\left( q^p+\e^2q^r\right)
-\frac12
\non \\
&+& \frac14 \beta^2 B^2 \left(pq^{p-2}+\e^2r q^{r-2}\right) 
+ 2\beta (B+T)\left[A(q)+\e^2 C(q)\right]
+\frac12 \beta^2 u B\left(pq^{p-1}+\e^2r q^{r-1}\right) \ ,
\eeqa
and where we used the following definitions,
\beqa
g(q) &=& -\frac{1}{2\beta}\log(1-q) 
 - \frac{\beta}{4}\left[ (p-1)q^p -pq^{p-1}+1\right]
\\
h(q) &=& 
- \frac{\beta}{4}\left[ (r-1)q^r -rq^{r-1}+1 \right]
\\
\frac{\partial g}{\partial m_i} &=& A(q) \, m_i
\\
\frac{\partial h}{\partial m_i} &=& C(q) \, m_i \ .
\eeqa
From (\ref{zizu}) we have that in the thermodynamic limit $N\to\infty$ 
the complexity is given by the effective action evaluated in the
saddle point values of the variational parameters $T,B,q,u$,
\beq
\Sigma(f) = \frac1N \log \overline{{\cal N}(f)} 
= \left. \hat S(T,B,q,u;\,f)\right|_{\rm saddle\ point} \ .
\eeq
The saddle point equations are not too difficult to solve. In fact, the parameters $T$ and $B$
can be worked out explicitly as functions of $q$. On the other hand, the equation for $q$ itself,
\beq
\frac{\partial \hat S(q,u;\, f)}{\partial q}= 0 \ ,
\label{qq}
\eeq
is highly nonlinear and must be solved numerically. 
The complexity is naturally a function of the free energy density $f$, so
in principle we have to minimize $\hat S$ also with respect to $u$, such that $f$ remains the only free
variable. However, $f$ and $u$ are conjugated variables (this is clear from the form of the action in
(\ref{action})). This means that we can either let $f$ free, work out $u=u(f)$ and study $\Sigma(f)$, or
let $u$ be the independent variable, such that $f=f(u)$ and study $\Sigma(u)$. It turns out that in
general $\Sigma$ is a smoother function of $u$ than $f$, so we will mostly use this last 
representation.

\begin{figure}
\includegraphics[clip,width=3 in]{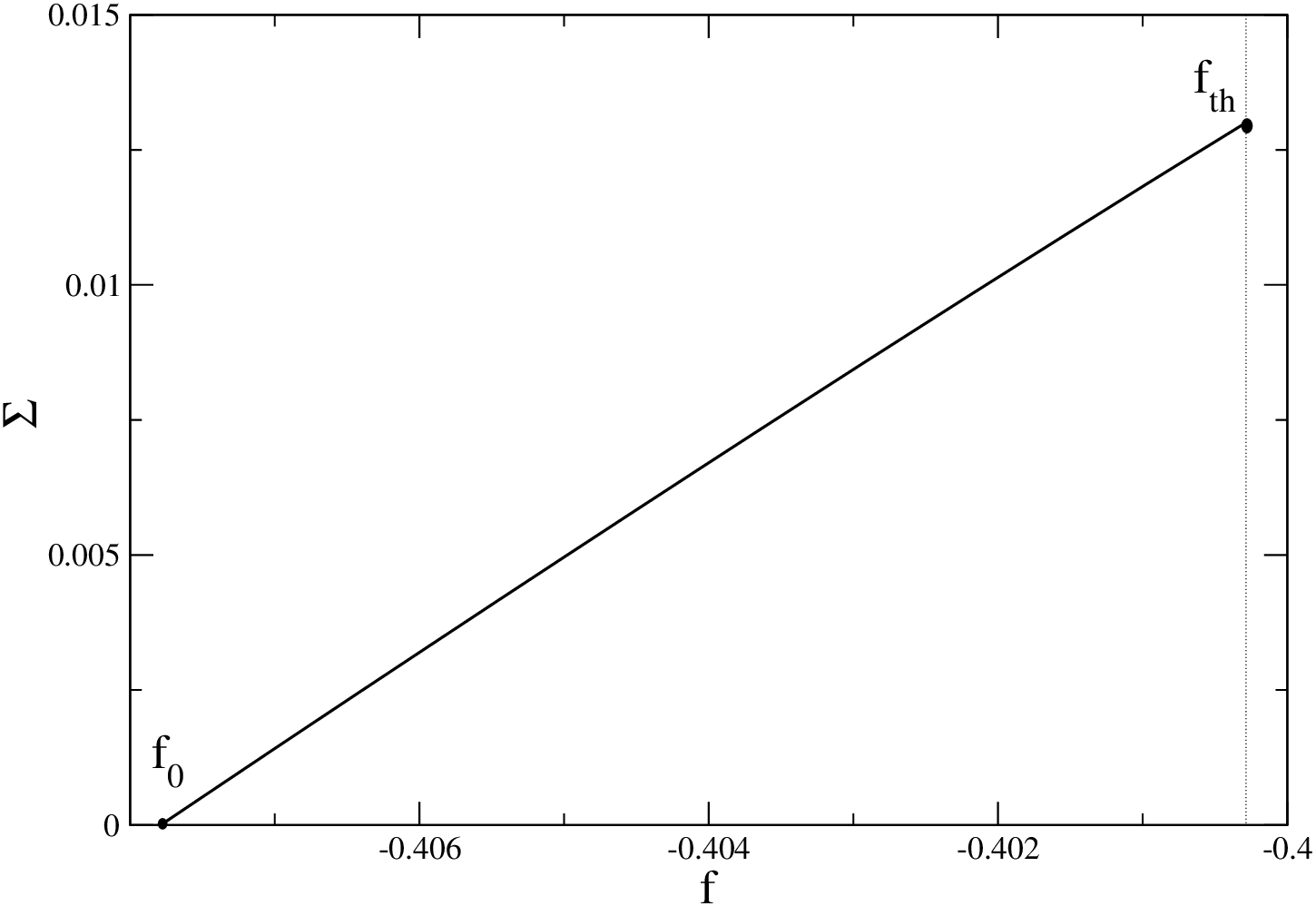}
\caption{Unperturbed $p$-spin spherical model: the complexity as a function of the free energy density.
The stability parameter $x_p$ is zero at the threshold $f=f_{th}$. The temperature is equal to the dynamical 
temperature $T_d=0.612$ of the unperturbed model.}
\label{fig0}
\end{figure}

Action (\ref{action}) is invariant under the generalized Becchi-Rouet-Stora-Tyutin (BRST) supersymmetry
\cite{brs,tito}-\cite{juanpe,noi1}. This symmetry generates the following two Ward identities  
\cite{juanpe,noi1},
\beq
\langle m \cdot x \rangle = - \langle \bar\psi \cdot \psi \rangle 
\quad\quad\quad  
\langle x \cdot x \rangle =  \langle u\, \bar\psi \cdot \psi \rangle \ .
\label{susy}
\eeq
The first Ward identity immediately translates into an equation for the parameters $T$ and $B$. From equations
(\ref{vara}) we have,
\beq
{\rm BSRT}\quad\Rightarrow\quad
B+T=0 \ .
\label{susina}
\eeq
Given that both $T$ and $B$ are functions of $q$, imposing this relation is equivalent to impose 
an extra equation for $q$, which has to be satisfied together with (\ref{qq}). By doing this we select
the supersymmetric solution of the saddle point equations. 
It is straightforward to prove that the second Ward identity in (\ref{susy}) gives a
relation for $q$ which is equivalent to (\ref{susina}).

\begin{figure}
\includegraphics[clip,width=3 in]{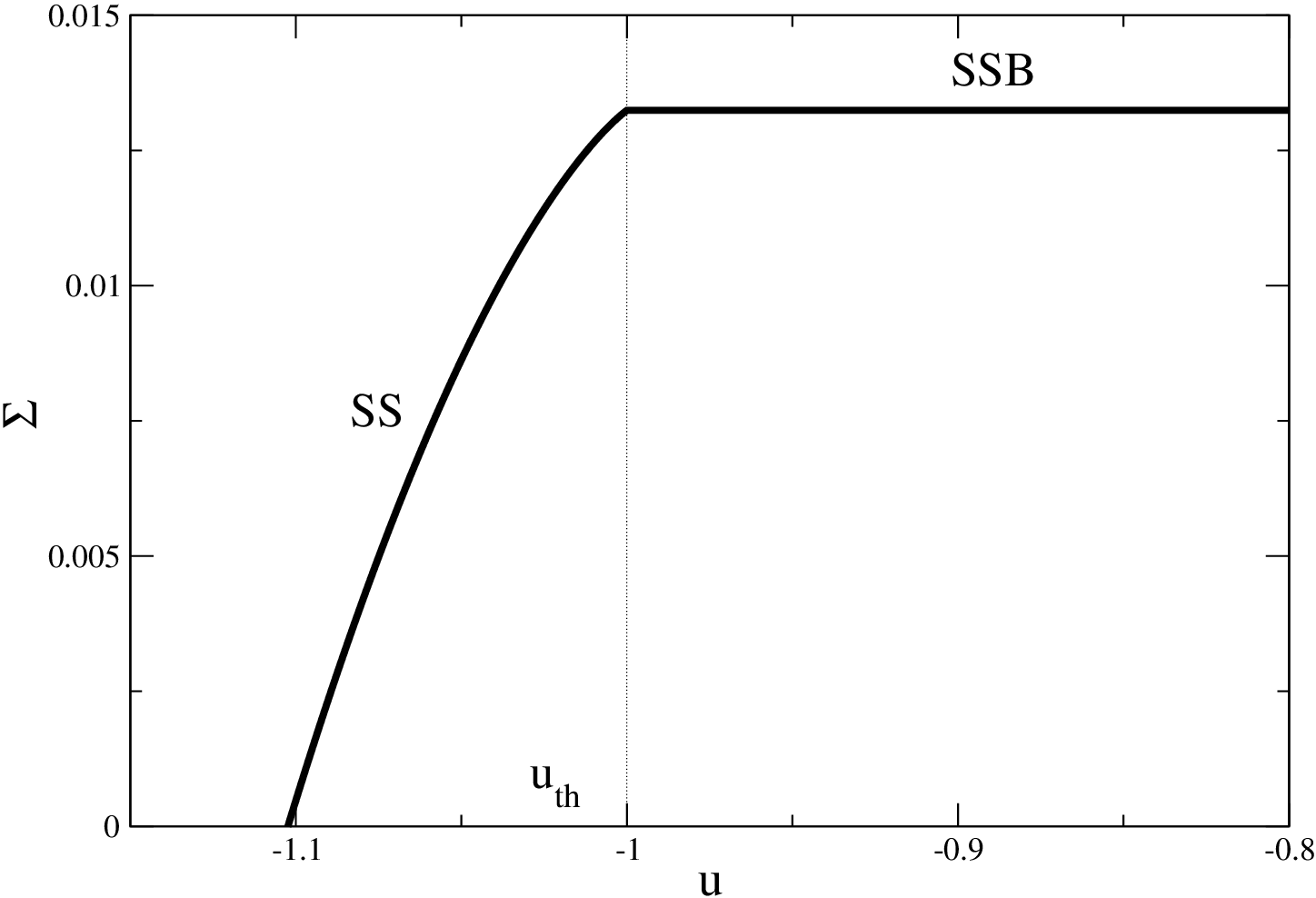}
\caption{Unperturbed $p$-spin spherical model. Complexity as a function of $u$. }
\label{fig1}
\includegraphics[clip,width=3 in]{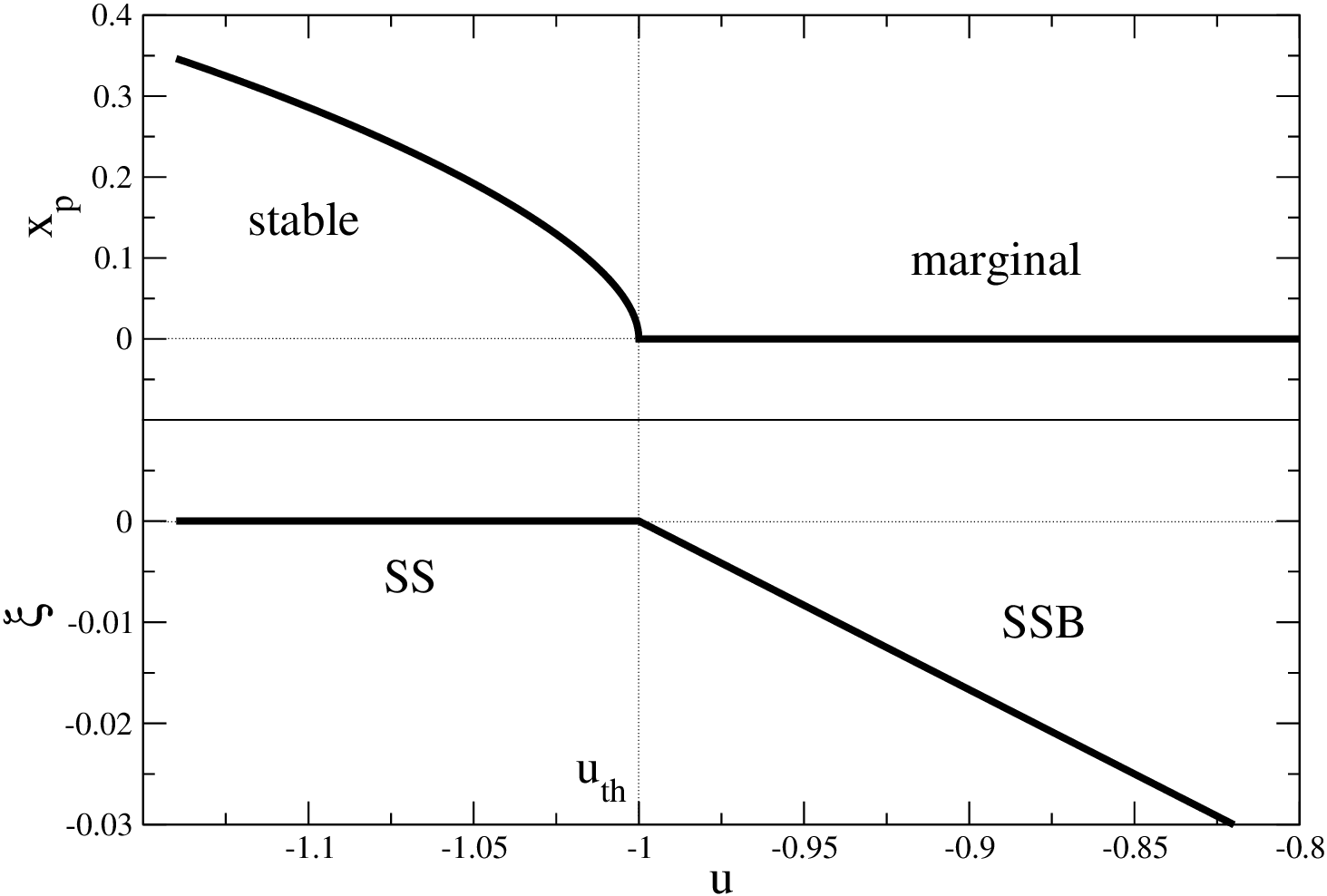}
\caption{Unperturbed $p$-spin spherical model. Upper panel: the stability parameter $x_p$. Lower panel:
the supersymmetry parameter $\xi$.}
\label{fig2}
\end{figure}

The saddle point equations for the complexity may, in general, have many solutions.
Some of these solutions are supersymmetric (SS), and therefore satisfy the Ward identities (\ref{susy}), whereas 
others are supersymmetry-breaking (SSB) solutions. These last solutions violate the Ward identities. Physically, different
solutions correspond to different structures of metastable states, so selecting the correct solution, 
whether supersymmetric or not, is a key point. As we have seen, from the studies 
performed so far, it seems that there are two classes of models. In 1RSB systems (as the unperturbed
$p$-spin spherical model) the supersymmetry is unbroken \cite{juanpe,leuzzi4}, so one can directly impose the Ward
identities (\ref{susy}) to simplify the saddle point equations and find the physical complexity. In the SK model
(and presumably in other FRSB systems), the supersymmetric saddle point is unstable, and physical 
metastable states are described by the SSB saddle point. In this case imposing 
(\ref{susy}) gives the wrong result.
In all these cases, the main way to discriminate the physical from the unphysical solution is to 
test the {\it stability} of the states described by the various solutions. In particular, a very
important quantity is the parameter introduced by Plefka in \cite{plefka},
\beq
x_p=1-\frac{\beta^2}{2N}\, p(p-1)q^{p-2}\,\sum_i(1-m_i^2)^2 \ .
\eeq
In order to have stable states, with a physical susceptibility, we must have,
\beq
x_p \ge 0 \ .
\label{stab}
\eeq

We shall use the stability condition (\ref{stab}) to classify physical and unphysical solutions
of the saddle point equation of our model. As a starting point we briefly discuss the case
$\e=0$, i.e. the unperturbed $p$-spin spherical model \cite{crisatap}. As we have said, in this case the physical
complexity is the one corresponding to the supersymmetric solution of the saddle point equations. Such solution 
can easily be found by imposing (\ref{susina}). The complexity $\Sigma$ as a function of
the free energy density $f$ (Fig.1)
is zero at the ground state $f_0$, and it is defined up to a {\it threshold} value $f_{th}$, beyond which 
the saddle-point equations have no longer solution. 
Threshold states are marginal, i.e. they have a zero mode in the second derivative of the free energy, 
and $x_p=0$ (whereas $x_p>0$ for all states with $f<f_{th}$).

\begin{figure}
\includegraphics[clip,width=3 in]{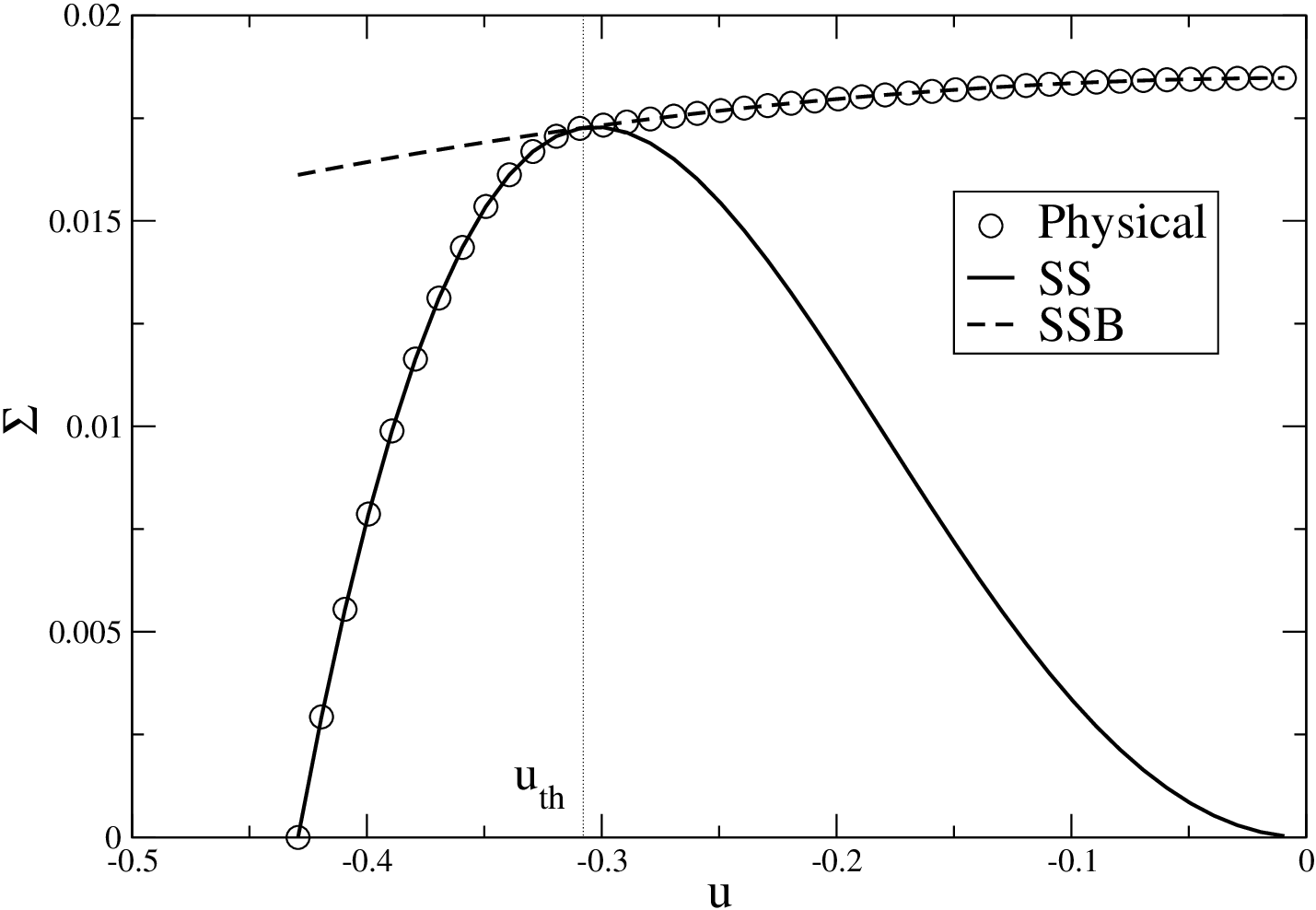}
\caption{Perturbed $p$-spin spherical model ($\e=0.45$). The complexity as a function of $u$. Full line: the SS
solution. Dashed line: the SSB solution. Open circles indicate the physical solution. $T=0.6 T_d$.}
\label{fig3}
\includegraphics[clip,width=3 in]{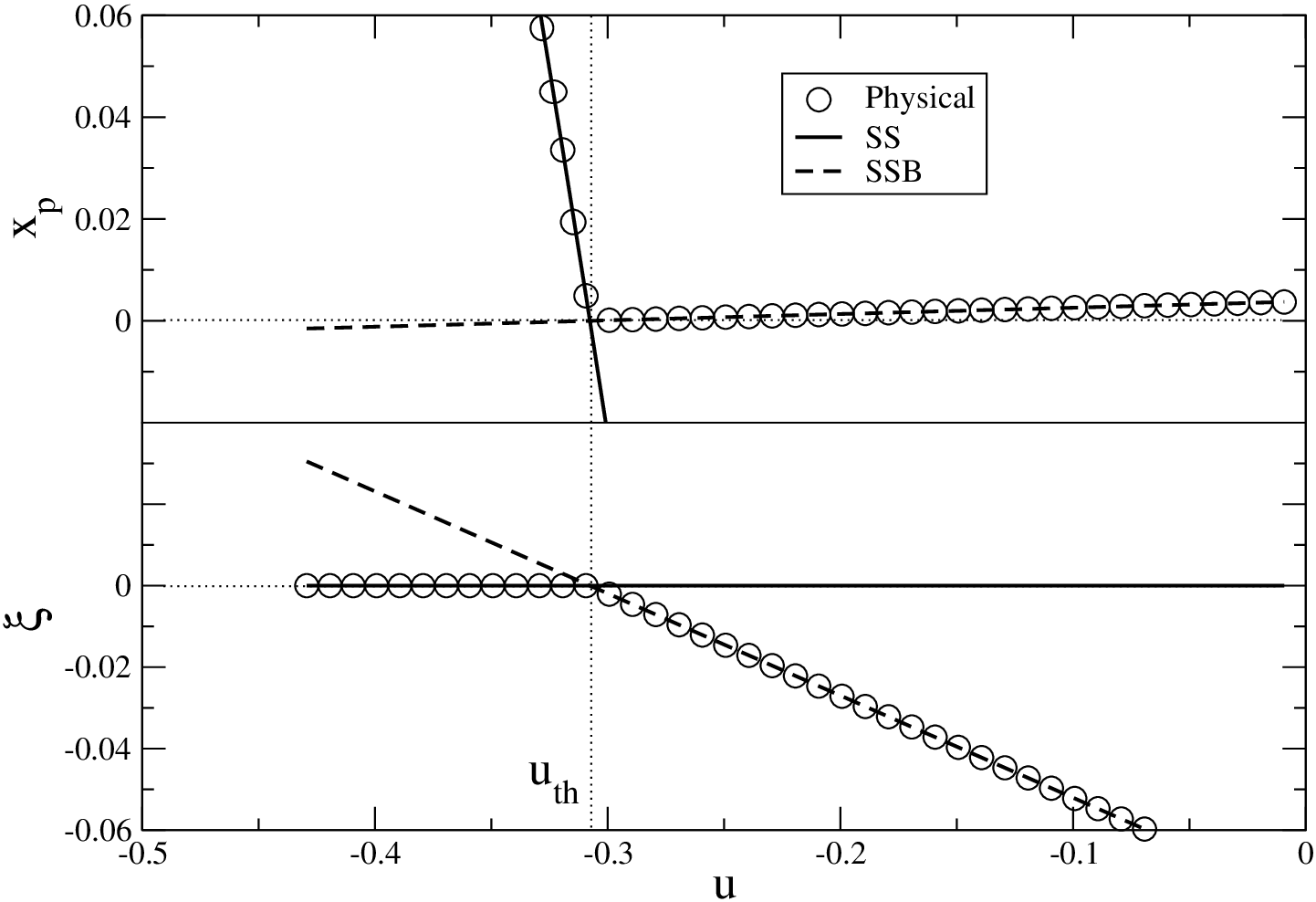}
\caption{Perturbed $p$-spin spherical model ($\e=0.45$). Upper panel: the stability parameter $x_p$. Lower panel:
the supersymmetry parameter $\xi$. $T=0.6 T_d$.}
\label{fig4}
\end{figure}

The fact that threshold states are marginal suggests, in analogy with the SK model \cite{abm,noi4},
that they may have a supersymmetry-breaking nature. If we study the complexity as a function of
the parameter $u$ (Fig.2), we find that $\Sigma(u)$ increases with $u$ up to a value $u_{th}=u(f_{th})$, and beyond
that point it remains constant, i.e. $\Sigma(u)=\Sigma_{th}$ for each $u\ge u_{th}$. Note that above $u_{th}$ 
also the free energy is constant, $f(u)=f_{th}$ for $u\geq u_{th}$, and thus the flat branch of $\Sigma(u)$ is
concentrated in the threshold point of $\Sigma(f)$ in Fig. 1.
The flat branch of the curve has $x_p=0$ (Fig.3), meaning, as we have already said, that threshold states are only
marginally stable. To assess the supersymmetric properties of this curve,  we can introduce the 
following parameter,
\beq
\xi=T+B \ .
\eeq
From (\ref{susina}) we have that $\xi=0$ on a SS saddle point, and $\xi\neq 0$
on a SSB saddle point. 
We see from Fig.3 that the flat branch of $\Sigma(u)$ has $\xi\neq 0$ and thus breaks the supersymmetry,
while the growing part of the curve is SS.
It should be noted that the flat branch continues up to $u=0$, which corresponds
to putting no constraint on the free energy, i.e. to counting the {\it total} number of states. In
other words, in the $p$-spin spherical model, the total, that is the maximum, 
complexity $\Sigma_{th}$  can be found either from the SS solution at $u=u_{th}$, or
from the SSB solution at $u=0$. The situation  in the unperturbed case seems therefore quite
strange: if we use $f$ as a natural variable, we have states up to $f=f_{th}$, and these are all supersymmetric
and stable. Only threshold states are marginally stable. On the other hand, using $u$ as a variable, we find
a degenerate interval between $u_{th}$ and $0$, where the complexity and the free energy remain stuck at 
their threshold levels, $f_{th}$ and $\Sigma_{th}$, and where the supersymmetry is broken. This situation
becomes clearer once we perturb the system.

When we switch on the perturbation $\e$ we recognize that the $\e=0$ case is just the peculiar
(and rather pathological) limit of a more general physical situation. 
We first analyze  $\Sigma$ as a function of $u$ (Fig.4). 
There are two solutions of the saddle point equations for each value of $u$:
the SS solution, and the SSB one. These two branches touch at
$u=u_{th}(\e)$. Below this point, the parameter $x_p$ is negative for the SSB solution, and positive 
for the SS one. On the other hand, $x_p$ changes sign for both solutions at $u_{th}$, and thus the 
physical stability of the two solutions switches at $u_{th}$. 
As a result SS states are the physical one below $u_{th}$, whereas 
SSB states become physical above $u_{th}$. 
As it can be seen from the plot, the SSB branch is very weakly dependent
on $u$ when $\e$ is small, and in the limit $\e=0$ it is just flat, as in Fig.2.

Therefore, threshold states are, in the general case, at the same time the highest SS 
and the lowest SSB states. This is in agreement with the marginal nature of
these states, as it was already well clear in the unperturbed case. It is interesting to note 
that the SSB solution in its stable phase $u>u_{th}$ is completely analogous to the SK model,
suggesting that the same connection holds also for the physical properties of the two models in 
this phase. The fact that the SK model has a FRSB thermodynamics, while the present one is a
strictly 1RSB model, is simply due to the fact that for the present model,
\beq
\Sigma(u_{th}) > 0 \ ,
\eeq
such that the SSB states do not contribute to the statics. The statics of the SK model is
FRSB simply because the threshold states in that model coincide with the ground states, i.e.
\beq
\Sigma(u_{th})=0 \quad\quad {\rm SK}\ .
\eeq
On the other hand, in the unperturbed $p$-spin model we have the opposite extreme behaviour:
there are no states whatsoever above the threshold, and therefore no SK phase, not even in the
metastable spectrum of states.

The crossover between a SS and a SSB phase can also be analyzed  by looking at the behaviour of 
$\Sigma$ as a function of $f$. Let us consider first the SS solution: the complexity grows from
$f_0$ where it is zero, up to a value $f_{th}=f(u_{th})$, where $u=u_{th}$. Beyond $u_{th}$ both $\Sigma$
and $f$ decreases, giving rise to a new branch of the SS curve $\Sigma(f)$. This backward branch
is SS, but unphysical, since $x_p<0$. The SSB solution takes over exactly at $f_{th}$, providing a
continuation of $\Sigma(f)$ beyond the threshold. This state of affairs is reproduced in 
Fig.6, where we plot $\Sigma$ in a neighborhood of $f_{th}$.
Therefore, unlike in the unperturbed $p$-spin spherical model, we have some physical states also above the 
threshold, and these states have a SSB structure.
Note that in this context  $f_{th}$ is not the largest free energy density of metastable states 
(as normally in the literature), but it is the {\it largest free energy density of supersymmetric 
metastable states}.

Our result suggests that the most general structure of metastable states in spin-glasses is the 
following. At low free energies lie SS states: these are non-marginal minima of the free energy, well separated from 
each other by large barriers. The stability of these states decreases with increasing energy: the 
softest mode of their second derivative matrix is smaller the larger $f$. At the threshold energy 
SS states become unstable, and above this point SSB states are the physical ones. Presumably, as in the
SK model, SSB states are marginal, in the sense that they have a zero mode leading out of the states,
and therefore are not well separated by large free energy barriers.

This crossover in free energy between SS and SSB states has important consequences on both the 
statical and dynamical behaviour of a system. As we have said above, when the transition free energy,
i.e. the threshold $f_{th}$ is {\it larger} than the ground state free energy $f_0$,
SS states dominate the thermodynamic properties, and the model is statically 1RSB. If, on the other hand,
$f_{th}$ is equal to $f_0$, the model has a FRSB statics. Note that the difference,
\beq
\Delta = f_{th}-f_0 \ ,
\eeq
may depend on the external parameters, as the temperature, the field or any other.
In the present model, by changing the temperature, $\Delta$ changes, but it always remains 
strictly positive, so that the model is always statically 1RSB. However, for other systems the situation 
may be different. In particular in the Ising $p$-spin model, there is a transition 
between a 1RSB and a FRSB phase at the Gardner temperature $T_G$ \cite{gardner}. Moreover, it has been proved in 
\cite{montanari} that in this model a crossover occurs from 1RSB to FRSB metastable states in the 
complexity.
Thus, what we expect is that in such a model the supersymmetric threshold becomes equal to the ground state at $T_G$,
i.e. $\Delta(T_G)=0$. The study of the supersymmetric properties of the Ising $p$-spin model,
will appear in \cite{neoleuzzi}. Preliminary results are in agreement with what we have shown here.

From a dynamical point of view, it is reasonable to assume that ergodicity is broken by the
highest non-marginal, i.e. supersymmetric, states. Therefore, we expect that the dynamical transition is
triggered by threshold states. In systems where $f_{th}>f_0$, as the spherical $p$-spin model,
we will have $T_d>T_s$, whereas in systems where $f_{th}= f_0$, as in the SK model, $T_d=T_s$.
We would like to remark that the high free energy transition between SS and SSB metastable states,
and its consequences on the static and dynamical behaviour of spin-glasses, was for the first time
discussed and understood in \cite{montanari}, albeit not in a supersymmetric context. In that study, 
the same phenomenon was interpreted as a 1RSB to FRSB transition in the spectrum of states. The physics, 
however, is the same.

\begin{figure}
\includegraphics[clip,width=3 in]{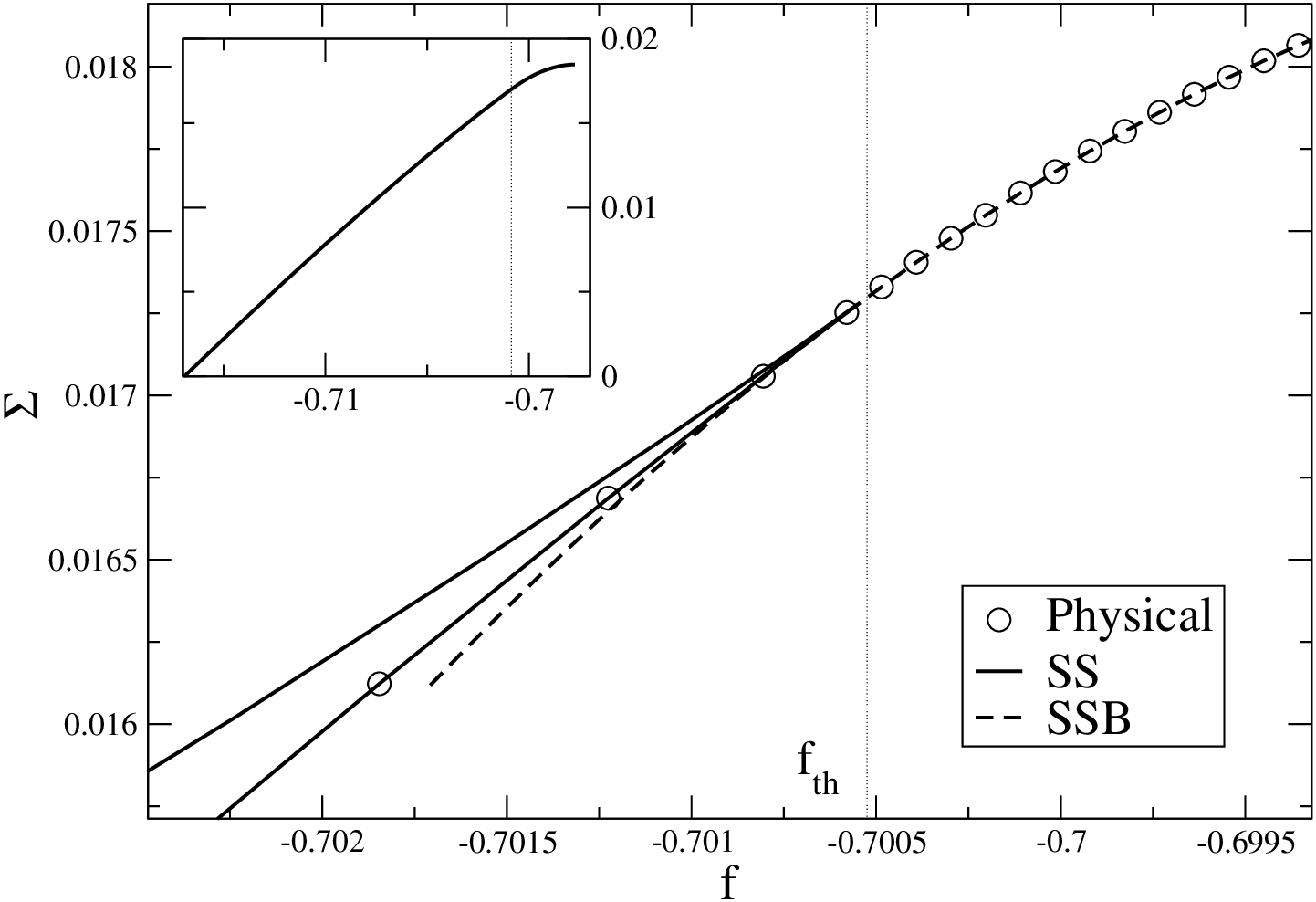}
\caption{Perturbed $p$-spin spherical model ($\e=0.45$). Complexity as a function of the free energy
density, around $f=f_{th}$. Inset: the full curve $\Sigma(f)$. $T=0.6 T_d$}
\label{fig5}
\end{figure}

In this paper we have analyzed a spin-glass model where a crossover takes place between low energies 
supersymmetric states, and higher energies supersymmetry-breaking states. The model is always
solved, at the static level,  by a 1RSB thermodynamic ansatz, although SSB states have the same physical 
structure as in FRSB systems. This is due to the fact that the crossover always occurs at energies higher 
than the ground state. Our study and the results of \cite{montanari} suggest 
that this may be the generic scenario, and that different static 
and dynamic behaviours may just depend on whether the transition free energy, i.e. the threshold, is larger 
or equal than the ground state free energy.

\begin{acknowledgments}
We warmly thank Irene Giardina, Luca Leuzzi, Giorgio Parisi and Federico Ricci-Tersenghi for numerous important discussions.
\end{acknowledgments}

\end{document}